\documentclass[aip,graphicx]{revtex4-1}
\usepackage{amssymb}
\usepackage{hyperref}
\usepackage{amsmath}
\usepackage{amsthm}
\usepackage{physics}
\usepackage[pdftex]{graphicx}
\usepackage{mathtools}
\usepackage{tikz}
\usepackage{color}

\newcommand*\circled[1]{\tikz[baseline=(char.base)]{
    \node[shape=circle,draw,inner sep=2pt] (char) {#1};}}

\draft 

\begin{document}


\title[Sobolev Spaces, Schwartz Spaces and \textcolor{black}{Electromagnetic and Gravitational} coupling]{Sobolev Spaces, Schwartz Spaces, and a definition of the Electromagnetic and Gravitational coupling }



\author{Jean-Philippe Montillet}
\affiliation{ESPlab, STI IMT, Ecole Polytechnique de Lausanne, MC A3 301 (Microcity), Rue de la Maladière 71b, CH-2002 Neuch$\hat{a}$tel, Switzerland}


\date{\today}

\begin{abstract}
The concept of \textit{multiplicity of solutions} was developed in \cite{JPMontillet2017} which is based on the theory of energy operators in the Schwartz space $\mathbf{S}^-(\mathbb{R})$ and some subspaces called energy spaces first defined in \cite{JPMontillet2013} and \cite{JPMontillet2014}. The main idea is to look for solutions of a given linear PDE in those subspaces. Here, this work extends previous developments in  $\mathbf{S}^-(\mathbb{R}^m)$ ($m \in \mathbb{Z}^+$) using the theory of Sobolev spaces, and in a special case the Hilbert spaces. Furthermore, we also define the concept of \textit{Energy Parallax}, which is the inclusion of additional solutions when varying the energy of a predefined system locally by taking into account additional smaller quantities. We show that it is equivalent to take into account solutions in other energy subspaces. To illustrate the theory, one of our examples is based on the variation of ElectroMagnetic (EM) energy density within the skin depth of a conductive material, leading to take into account derivatives of EM evanescent waves, particular solutions of the wave equation. The last example is the derivation of the Woodward effect \cite{Woodward} with the variations of the EM energy density under strict assumptions in general relativity. It finally leads to a theoretical definition of an electromagnetic and gravitational \textcolor{black}{(EMG)} coupling.
\end{abstract}


\maketitle 
\newtheorem{theorem}{Theorem}[section]
\newtheorem{proposition}[theorem]{Proposition}
\newtheorem{corollary}[theorem]{Corollary}


\section{Overview}
Teager-Kaiser energy operator was defined in \cite{Kaiser90} and the family of Teager-Kaiser energy operators in \cite{Maragos1995}. Many applications in signal processing were found over the past $25$ years such as detecting transient signals \cite{Dunn}, filtering modulated signals \cite{Bovik93}, image processing \cite{Cexus} and in localization \cite{Schasse}. However, \cite{JPMontillet2010} and \cite{JPMontillet2013} introduced the conjugate Teager-Kaiser energy operator and associated family $( \Psi_k^+)_{k\in\mathbb{Z}}$. Subsequently using iterations of the Lie Bracket, \cite{JPMontillet2014} defined the generalized conjugate Teager-Kaiser energy operators $( [[.]^p]_k^+)_{k\in\mathbb{Z}}$ ($p \in \mathbb{Z}^+$). To abbreviate the notation, we sometimes use the generic name  \textit{energy operator} in order to refer to the conjugate Teager-Kaiser energy operators and the generalized conjugate Teager-Kaiser energy operators. Precision is made in the denomination when it is required. Furthermore, the purpose of the energy operators and generalized energy operators was the decomposition of the successive derivatives of a finite energy function $f^n$ ($n$ in $\mathbb{Z}^+ -\{0,1\}$) in the Schwartz space $\mathbf{S}^-(\mathbb{R})$ . The generalized energy operators were introduced when decomposing the successive derivatives of a finite energy function of the form $([[f]^p]_1^+)^n$ ($n$ in $\mathbb{Z}^+ -\{0,1\}$) in the Schwartz space.
It then follows in \cite{JPMontillet2014} and \cite{JPMontillet2017} the definition of \textit{Energy Spaces}, which are subspaces of the Schwartz Space $\mathbf{S}^-(\mathbb{R})$ associated with energy operators and generalized energy operators. This definition was used to define the concept of \textit{multiplicity of solutions} in \cite{JPMontillet2017} ($Theorem$ $2$ and $Corollary$ $1$).  The idea is to consider those energy spaces and functions associated with them when solving linear PDEs. More precisely, we look for solutions of a nominated linear PDE within those energy spaces (including the space reduced to $\{0\}$). The concept was further developed using the Taylor series of the energy of a solution $\mathbf{S}^-(\mathbb{R})$  for a nominated PDE. The work was based on finding when the successive derivatives, defined through the Taylor series coefficients, are also solutions of this particular PDE (see Section $4$ in \cite{JPMontillet2017}). 
\\ This work first generalizes in $\mathbf{S}^-(\mathbb{R}^m)$ ($m$ $\in \mathbb{Z}^+$) the theorems and lemmas established in \cite{JPMontillet2013} and \cite{JPMontillet2014} stated for $\mathbf{S}^-(\mathbb{R})$  using the properties of the $L^2$ space called here $L^2(\mathbb{R}^m)$ ($m$ $\in \mathbb{Z}^+$) together with the general property of the Schwartz space $\mathbf{S}^-(\mathbb{R}^m) \subset L^2(\mathbb{R}^m)$ ($m$ $\in \mathbb{Z}^+$) \cite{Schwartz}. However, this work imposes the condition of the stability by Fourier transform for any functions in $\mathbf{S}^-(\mathbb{R})$ in order to use the Sobolev space(see $\bold{Appendix}$ $I$, $\mathbf{Definition}$ $I.1$). Thus, in this work we consider  $\mathbf{S}^-(\mathbb{R})$ together with its dual: the tempered distributions $\mathbf{S}^{*,-}(\mathbb{R})$.
 Secondly, the energy spaces $\mathbf{M}_p^k$ ($p$ $\in \mathbb{Z}^+$, $k$ $\in \mathbb{Z}^+$) are also redefined as subspaces of $\mathbf{S}^-(\mathbb{R}^m)$. Furthermore, with the definition of the Sobolev spaces, and in particular the Hilbert spaces $\mathbf{H}^k(\mathbb{R}^m)$, it allows to  show the inclusion $\mathbf{M}_p^k \subset \mathbf{H}^k(\mathbb{R}^m)  \subset L^2(\mathbb{R}^m)$. Then, we finally redefine in $\mathbb{R}^m$ the $Theorem$ $3$ established in \cite{JPMontillet2017} and the concept of \textit{multiplicity of solutions}. 
\\ The first section together with $\bold{Appendix}$ $I$ are reminders about some important definitions and properties for the Sobolev spaces, the Schwartz space and the L$2$-norm. Section \ref{SectionrefSobolevAll} deals with the generalization of the work exposed in \cite{JPMontillet2013} and \cite{JPMontillet2014} in $\mathbf{S}^-(\mathbb{R}^m)$ and the redefinition of the energy spaces. Section \ref{Multiplicity} recalls  the concept of \textit{multiplicity of solutions} defined in \cite{JPMontillet2017} and generalized in $\mathbf{S}^-(\mathbb{R}^m)$ with $\mathbf{Theorem}$ $4$.  The last section focuses on some applications of this theory. The first application is the wave equation and the discussion of taking into account more solutions from other energy spaces. We then define another concept called \textit{energy parallax} (i.e. mathematically in $\bold{Definition}$ $4$, see discussion on the  physical interpretation in $\bold{Appendix}$ $II$) which is directly related to \textit{multiplicity of solutions}. In order to illustrate this concept, a second example is the variation of energy density in the skin depth of a conductor material. The idea is to show that the variation of energy density can lead to consider multiple derivatives of evanescent waves resulting from the electromagnetic field. The last section is dedicated to the derivation of the Woodward effect \cite{Woodward}from the Hoyle-Narlikar theory \cite{Heidi, Woodwardb} using the EM energy density and a discussion takes place about the relationship to the presented theory of energy spaces. \textcolor{black}{It leads to a theoretical definition of an Electromagnetic and Gravitational coupling (EMG).} 
\section{ Definition of L-2 norm and Schwartz space}
\textcolor{black}{
\subsection{Notation and Symbols}
In this work, several symbols are used. The set of integer numbers $\mathbb{Z}$ is sometimes called only for the positive integer such as $\mathbb{Z}^+$ or $\mathbb{Z}^m_+$ (for a space with dimension $m$). When the integer $0$ is not included, it is explicitly mentioned such as   $\mathbb{Z}^+-\{0\}$. The set of natural numbers is $\mathbb{N}$, with only the positive numbers defined as  $\mathbb{N}^+$. $\mathbb{R}$ is the set of real numbers.  Also, the Schwartz space is here called $\mathbf{S}^{-}(\mathbb{R}^m)$ which is the notation used in previous works such as \cite{JPMontillet2017} and \cite{JPMontillet2013}. Several notations describe the relationship between spaces such as intersection ($\bigcap$), union ($\bigcup$), inclusion ($\subset$, inclusion without the equality $\subsetneq$,  inclusion with equality $\subseteq$). Reader can refer to \cite{SobolevB} or advanced mathematical textbooks for more explanations.}
\textcolor{black}{
\subsection{ L-2 norm and Schwartz space}
}
With the difference in  $\bold{Appendix}$ $I$ and the generalities with the Sobolev spaces, here the analysis focuses on the L-$2$ norm ($p$ equal to $2$ for the $L^p$ norm). It allows to state the Plancherel identity $\forall$ $f$ $\in L^2(\mathbb{R}^m)$ :
\begin{equation}
\int_{R^m} \abs{f}^2 dt = \int_{R^m} \abs{\mathcal{F}(f)(\xi)}^2 d\xi 
\end{equation}
We are here interested in the functions belonging to the Schwartz space $f$ $\in$ $\mathbf{S}^-(\mathbb{R}^m)$ $\subset  L^2(\mathbb{R}^m)$. \textcolor{black}{The Schwartz space consists of smooth functions whose derivatives (including the function)
are rapidly decreasing (e.g., the space of all bump functions \cite{stevenG})}. The Schwartz space $\mathbf{S}^{-}(\mathbb{R}^m)$ is defined as ( for $m$ $\in$ $[1,2]$ \cite{JPMontillet2017, JPMontillet2014}, for $m$ $\in$ $\mathbb{Z}^+$  \cite{SobolevB, PaperUkn}:
\begin{equation}\label{SRRRRR}
\mathbf{S}^{-}(\mathbb{R}^m) =\{f \in \mathbf{C}^{\infty}(\mathbb{R}^m)| \norm{f}_{\alpha, \beta} < \infty, \qquad \textcolor{black}{\forall} \alpha, \beta \in \mathbb{Z}^m_+  \}
\end{equation}
where $\alpha$, $\beta$ are  multi-indices and  
\begin{equation}
\norm{f}_{\alpha, \beta} = sup_{t \in \mathbb{R}^m} |t^\beta D^\alpha f(t)|
\end{equation}
Note that one can define $\mathbf{S}^{-}(\mathbb{R}^m)$ with $\textcolor{black}{\forall \alpha, \beta \in \mathbb{Z}^m}$ according to \cite{Folland}, but we decide to use  $\mathbb{Z}^m_+$ following the development in the next sections. It is useful for the remainder of the work to remember some properties of the Schwartz functions in $\mathbf{S}^-(\mathbb{R}^m)$.
\vspace{0.5em}
\\ $\bold{\textcolor{black}{Properties}}$ $1$ \cite{UCDavis}: Some Properties of  $\mathbf{S}^-(\mathbb{R}^m)$.
\begin{itemize}
\item If $1 \leq  p \leq \infty$, then $\mathbf{S}^-(\mathbb{R}^m)\subset L^p(\mathbb{R}^m)$
\item   $\mathbf{S}^-(\mathbb{R}^m)$ is a dense subspace of ${H}^{k,2}(\mathbb{R}^m)$ ( $k \in \mathbb{N}$).
\item (Stability with Fourier transform) The Fourier transform is a linear isomorphism $\mathbf{S}^-(\mathbb{R}^m)\rightarrow \mathbf{S}^-(\mathbb{R}^m)$.
\item If $f$ $\in$ $\mathbf{S}^-(\mathbb{R}^m)$, then $f$ is uniformly continuous on $\mathbb{R}^m$.
\end{itemize}
The proof of those properties are standard results with Schwartz spaces established in many harmonic analysis books (e.g., \cite{Folland}, \cite{UCDavis}).
\vspace{0.5em}
\\ $\bold{Remark}$ $(1)$  Note that in \cite{JPMontillet2017}, \cite{JPMontillet2013}, \cite{JPMontillet2014}, the author used the general term of \textit{finite energy functions} for Schwartz functions in $\mathbf{S}^-(\mathbb{R}^m)$, with $m$ restricted to $[1,2]$. It is a common definition in signal processing for the functions in $L^2(\mathbb{R}^m)$ and generally associated with the Plancherel identity.
\vspace{0.5em}
\\ $\bold{Remark}$ $(2)$ One way to interpret the property that $\mathbf{S}^-(\mathbb{R}^m)$ is stable by Fourier transform is:

  for $f$ $\in$ $\mathbf{S}^-(\mathbb{R}^m)$, $k$ $\in$ $\mathbb{N}$
\begin{eqnarray}
sup_{\xi \in \mathbb{R}^m} \{|(1+\abs{\xi}^2)^{k/2}\mathcal{F}(f)(\xi)|\} &<& \infty  \nonumber \\
\leftrightarrow \hspace{0.5em} \exists a \in \mathbb{R}, \hspace{0.5em} |(1+\abs{\xi}^2)^{k/2}\mathcal{F}(f)(\xi)| & \leq & \frac{a}{1+\abs{\xi}^2}
\end{eqnarray}
Now, let us recall the definition of the Hilbert spaces ${H}^{k,p}(\mathbb{R}^m)$ (Sobolev spaces ${W}^{k,p}(\mathbb{R}^m)$ for $p=2$, see $\bold{Appendix}$ $I$, $\bold{Definition}$ $I.1$) from \eqref{HilbertSobolevdef} and drop the sup-script $p$ in the remainder of this work:
\begin{equation}\label{HilbertSobolevdef2}
{W}^{k}(\mathbb{R}^m) = {H}^{k}(\mathbb{R}^m) := \{f \in \mathbf{S}^{*,-}(\mathbb{R}^m)\lvert  \big (1+ \abs{\xi}^2)^{k/2}\mathcal{F}(f)  \in L^2(\mathbb{R}^m) \}
\end{equation}
\textcolor{black}{Note that $\mathbf{S}^{*,-}(\mathbb{R}^m)$ is the space of tempered distributions, dual of $\mathbf{S}^-(\mathbb{R}^m)$ via the Fourier transform.}  A function belongs to $L^2(\mathbb{R}^m)$ if and only if its Fourier transform belongs to $L^2(\mathbb{R}^m)$ and the Fourier transform preserves the $L^2$-norm. As a result, the Fourier transform provides a simple way to define $L^2$-Sobolev spaces on $\mathbb{R}^m$ (including ones of fractional and negative order $m$ \cite{UCDavis}). \textcolor{black}{Finally, the stability via Fourier transform is the key for $\mathbf{S}^-(\mathbb{R}^m) \subsetneq \mathbf{H}^{k}(\mathbb{R}^m)$.} 
\vspace{0.5em}
\\ $\bold{Remark}$ $(3)$ Following the remark (\textit{Remark} $3.4$ in \cite{SobolevA} ) and the general properties of the Fourier transform, one can state the equivalence relationship in $L^2(\mathbb{R}^m)$
\begin{eqnarray}
f \in \mathbf{H}^{k}(\mathbb{R}^m) \leftrightarrow D^{\alpha} f \in L^2(\mathbb{R}^m) \forall \abs{\alpha} \leq k &\leftrightarrow & \mathcal{F}(D^{\alpha} f) \in L^2(\mathbb{R}^m) \forall \abs{\alpha} \leq k  \nonumber \\
\leftrightarrow \xi^\alpha \mathcal{F}(f) \in L^2(\mathbb{R}^m) \forall \abs{\alpha} \leq k &\leftrightarrow&  (1+\abs{\xi}^2)^{\abs{\alpha}/2}\mathcal{F}(D^{\alpha} f) \in L^2(\mathbb{R}^m) \forall \abs{\alpha} \leq k  \nonumber
\end{eqnarray}
Using the definition of $\mathbf{H}^{k}(\mathbb{R}^m)$ and the properties of the Fourier transform, it is also possible to show that for $k>k'$, $\mathbf{H}^{k}(\mathbb{R}^m) \subset \mathbf{H}^{k'}(\mathbb{R}^m)$ \cite{aman}, and the relationship $\mathbf{H}^{0}(\mathbb{R}^m) = L^2(\mathbb{R}^m) $. 
 It is also possible to define $\mathbf{H}^{\infty}(\mathbb{R}^m) = \bigcap_{k \in \mathbb{N}} \mathbf{H}^{k}(\mathbb{R}^m)$ with $\mathbf{S}^-(\mathbb{R}^m) \subset \mathbf{H}^{\infty}(\mathbb{R}^m)$, and to extend this equality to $k$ $\in \mathbb{R}$ following \cite{SobolevA}.
\section{On Some Subsets of Schwartz Spaces: Energy Spaces}\label{SectionrefSobolevAll}
This section first recalls generalities on the Teager-Kaiser energy operator and its conjugate operator with the application to decompose Schwartz functions from the work developed in \cite{JPMontillet2013} and \cite{JPMontillet2014}. We call in this work \textit{Energy operators} the families of operators based on the Teager-Kaiser energy operator. The definitions and theorems are here stated for the Schwartz space $\mathbf{S}^-(\mathbb{R}^m)$ ($m \in \mathbb{N}$) whereas the preliminary work in \cite{JPMontillet2013} and \cite{JPMontillet2014} stated the definitions and main theorems for $m$ $\in$ $[1,2]$. For $m=2$ in Section $6$ in \cite{JPMontillet2014}, a discussion takes place during the application of the theory to linear partial differential equations. Secondly, the energy spaces defined in \cite{JPMontillet2017} and \cite{JPMontillet2014} are here generalized on $\mathbf{S}^-(\mathbb{R}^m)$ with novel relationships with Sobolev spaces $\mathbf{H}^{k}(\mathbb{R}^m)$ ($k \in \mathbb{N}$).
\subsection{Definition and properties of the Energy operators in $\mathbf{S}^-(\mathbb{R}^m)$ }\label{SectionrefSobolev}
Let us call the set $\mathcal{F}(\mathbf{S}^{-}(\mathbb{R}^m),\mathbf{S}^{-}(\mathbb{R}^m))$ all Schwartz functions (or operators) defined such as $\gamma:$ $\mathbf{S}^{-}(\mathbb{R}^m)$ $\rightarrow$ $\mathbf{S}^{-}(\mathbb{R}^m)$. 
For $f$ $\in$ $\mathbf{S}^{-}(\mathbb{R}^m)$, let us define $\mathbf{\partial}_i^k f$ ($k$ $\in \mathbb{Z}$, $i$ $\in [1,...,m]$), with $f$ defined with the vector parameter  $\mathbf{T} =[t_1, t_2, ...,t_m]$ $\in$ $\mathbb{R}^m$ such as
\begin{equation}\label{multiindexOp}
\left\{
\begin{array}{lr}
\partial_i^k f = &\frac{\partial^k f}{\partial t_i^k}, \hspace{0.5em} \forall i \in [1,...,m], \hspace{0.5em} \forall k \in \mathbb{Z}^+-\{0\}  \\
\partial_i^k f = & \int_{-\infty}^{t_i} \big (... \big ( \int_{-\infty}^{\tau_1} f(t_1,t_2,..,\tau_1,t_{k+1},t_m) d\tau_1 \big ) ...\big ) d\tau_k   , \hspace{0.5em} \forall i \in [1,...,m], \hspace{0.5em} \forall k \in \mathbb{Z}^- -\{0\} \\
\partial_i^0 f = & f , \hspace{0.5em} \forall i \in [1,...,m]
\end{array} \right. 
\end{equation}
Combining multiple integrals and derivatives justify the use of the Schwartz space $\mathbf{S}^{-}(\mathbb{R}^m)$ and echoes the choice made previously in \cite{JPMontillet2013} (see equation ($10$)).
%
The definitions and results given in \cite{JPMontillet2013} and \cite{JPMontillet2014} in the case $\mathbf{S}^{-}(\mathbb{R})$  are now formulated for $\mathbf{S}^{-}(\mathbb{R}^m)$.
Section $2$ in \cite{JPMontillet2013} and Section $4$ in \cite{JPMontillet2014} defined the energy operators $\Psi_k^+$, $\Psi_k^-$ ($k$ in $\mathbb{Z}$) and the generalized energy operators $[[.]^p]_k^+$ and $[[.]^p]_k^-$ ($p$ in $\mathbb{Z}^+$). Following \cite{JPMontillet2014}, let us define the energy operators with multi-index derivative in \eqref{multiindexOp}:
\begin{eqnarray}\label{bracketl}
\Psi_k^+(.) &=&  \sum_{i=1}^m \partial_i^1.\partial_i^{k-1}.+ \partial_i^{0}. \partial_i^{k}. \nonumber\\
\Psi_k^+(.) &=& \sum_{i=1}^m \psi_{k,i}^+(.) \nonumber\\
{[.,.]}_k^+ &=& \Psi_k^+(.) \nonumber\\
{[.,.]}_{k,i}^+ &=& \psi_{k,i}^+(.) \nonumber\\
\end{eqnarray}
Further more, we also use the short notation ${[.,.]}_k^+ = [.]_k^+ $ in the remainder of this work. Note that $\Psi_k^-$ is the conjugate operator of $\Psi_k^+$ and $\psi_{k,i}^-$ respectively to $\psi_{k,i}^+$.
\vspace{0.5em}
\\ $\bold{Remark}$ $(4)$ The families of (generalized) energy operators  $([[.]^p]_k^+)_{k\in\mathbb{Z}} $ and $([[.]^p]_k^-)_{k\in\mathbb{Z}} $ ($p$ in $\mathbb{Z}^+$) are also called families of differential energy operator (DEO) \cite{JPMontillet2013} \cite{JPMontillet2014}.
\vspace{0.5em}

 Furthermore, \cite{JPMontillet2014} defined the generalized energy operators $[[.]^1]_{k}^+$ and $[[.]^1]_{k}^-$ ($k \in \mathbb{Z}$):
\begin{eqnarray}\label{generalizedEO}
{[[.,.]_{k,i}^+,[.,.]_{k,i}^+]}_{k,i}^+ &=& \partial_i^1 {\psi}_{k,i}^+(.) \partial_i^{k-1} {\psi}_{k,i}^+(.) +  \partial_i^0{\psi}_{k,i}^+(.) \partial_i^k {\psi}_{k,i}^+(.) \nonumber \\
{[[.,.]_{k,i}^+,[.,.]_{k,i}^+]}_{k,i}^+ &=& \partial_i^1 [[.]^0]_{k,i}^+ \partial_i^{k-1} [[.]^0]_{k,i}^+ +  \partial_i^0[[.]^0]_{k,i}^+ \partial_i^k [[.]^0]_{k,i}^+ \nonumber \\
{[[.,.]_k^+,[.,.]_k^+]}_{k}^+ &=&  \sum_{i=1}^m {[[.,.]_{k,i}^+,[.,.]_{k,i}^+]}_{k,i}^+ \nonumber \\
{[[.,.]_k^+,[.,.]_k^+]}_{k}^+ &=&  \sum_{i=1}^m [[.]^1]_{k,i}^+ \nonumber \\
&=&[[.]^1]_{k}^+ \nonumber \\
\end{eqnarray} 
By $\it{iterating}$ the bracket $[.]$, \cite{JPMontillet2014} defined the generalized operator  $[[.]^p]_{k,i}^-$ and the conjugate $[[.]^p]_{k,i}^+$ with $p$ in $\mathbb{Z}^+$. Note that $[[f]^p]_{1,i}^- =0$ $\forall$ $p$ in $\mathbb{Z}^+$ and $i$ in  $ \mathbb{Z}$.
\newline Now, the derivative chain rule property and bilinearity of the energy operators and generalized operators (for $i$ in $[1,2]$) are shown respectively in\cite{JPMontillet2013}, Section $2$ and \cite{JPMontillet2014}, Proposition $3$. The generalisation of this property to $i$ in $[1,..,m]$ for the operators $\psi_{k,i}^+(.)$, $\psi_{k,i}^-(.)$, $[[.]^p]_{k,i}^-$ and $[[.]^p]_{k,i}^+$ ($k$  $\in \mathbb{Z}$, $p$ $\in\mathbb{Z}^+$) is trivial due to the linearity of the derivatives and integrals when defining $\partial_i^k$ in \eqref{multiindexOp}. Due to the linearity of the sum, the bilinearity property is also generalized to $ \Psi_k^+(.)$, $ \Psi_k^-(.)$, $[[.]^1]_{k}^+$ and $[[.]^1]_{k}^-$ ($k$  $\in \mathbb{Z}$, $p$ $\in\mathbb{Z}^+$). 
\vspace{1.0em}
\\$\bold{Definition}$ $1$ \textit{\cite{JPMontillet2013}}: $\forall$ $f$ in $\mathbf{S}^{-}(\mathbb{R}^m)$, $\forall$ $v\in\mathbb{Z}^+-\{0\}$, $\forall$  $n\in\mathbb{Z}^+$ and $n>1$, the family of operators $(G_k)_{k \in \mathbb{Z}}$ (with $(G_k)_{k \in \mathbb{Z}}$ $\subseteq$ $\mathcal{F}(\mathbf{S}^{-}(\mathbb{R}^m),\mathbf{S}^{-}(\mathbb{R}^m))$) decomposes $\partial_i^v$$f^n$ in $\mathbb{R}^m$ ($i$ $\in$ $[1,...,m]$), if it exists $(N_j)_{j\in \mathbb{Z}^+ \cup \{0\}}$ $\subseteq$ $\mathbb{Z^+}$,  $(C_l)_{l=-N_j}^{N_j}$ $\subseteq$ $\mathbb{R}$, and it exists $(\alpha_j)$ and $r$ in $\mathbb{Z^+}\cup\{0\}$ (with $r<v$) such as $\partial_i^v$$f^n$ $=$ $\sum_{j=0}^{v-1} \big(_{j}^{v-1} \big) \partial_i^{v-1-j} f^{n-r}$ $\sum_{u=-N_j}^{N_j} C_u G_u(\partial_i^{\alpha_u}f)$.
\vspace{1.0em}
\\ In addition, one has to define $\mathbf{s}^{-}(\mathbb{R}^m)$ as:
\begin{equation}
\mathbf{s}^{-}(\mathbb{R}^m) = \{ f \in  \mathbf{S}^{-}(\mathbb{R}^m) | f \notin  (\cup_{k \in \mathbb{Z}} Ker(\Psi^{+}_k))\cup(\cup_{k \in \mathbb{Z}-\{1\}} Ker(\Psi^{-}_k))\}
\end{equation}
or with the energy operators $\psi_{k,i}^+$ and $\psi_{k,i}^-$ defined in \eqref{bracketl}
\begin{equation}
\mathbf{s}^{-}(\mathbb{R}^m) = \{ f \in  \mathbf{S}^{-}(\mathbb{R}^m) | f \notin  \cup_{i\in [1,...,m]} \big ( \cup_{k \in \mathbb{Z}} Ker(\psi_{k,i}^+(f))\cup(\cup_{k \in \mathbb{Z}-\{1\}} Ker(\psi_{k,i}^-(f)) \big)\}
\end{equation}
$Ker(.)$ is the notation for the kernel associated here with the operators $\Psi^{+}_k$, $\Psi^{-}_k$,  $\psi_{k,i}^+$ and  $\psi_{k,i}^-$ ($k$ in $\mathbb{Z}$) (see \cite{JPMontillet2013}, $Properties$ $1$ and $2$). By definition, one can state that $\mathbf{s}^{-}(\mathbb{R}^m) \subsetneq \mathbf{S}^{-}(\mathbb{R}^m)$.    Following $\bold{Definition}$ $1$, the \emph{uniqueness} of the decomposition  in $\mathbf{s}^{-}(\mathbb{R}^m)$ with the families of differential operators can be stated as:
\vspace{1.0em}
\newline $\bold{Definition}$ $2$ \textit{\cite{JPMontillet2013}}: $\forall$ $f$ in $\mathbf{s}^{-}(\mathbb{R}^m)$, $\forall$ $v\in\mathbb{Z}^+-\{0\}$, $\forall$  $n\in\mathbb{Z}^+$ and $n>1$, the families of operators $(\Psi^{+}_k)_{k \in \mathbb{Z}}$ and $(\Psi^{-}_k)_{k \in \mathbb{Z}}$ ($(\Psi^{+}_k)_{k \in \mathbb{Z}}$ and $(\Psi^{-}_k)_{k \in \mathbb{Z}}$ $\subseteq$ $\mathcal{F}(\mathbf{s}^{-}(\mathbb{R}^m),\mathbf{S}^{-}(\mathbb{R}^m)$) decompose uniquely $\partial_i^v$ $f^n$ in $\mathbb{R}^m$, if for any family of operators $(S_k)_{k \in \mathbb{Z}}$ $\subseteq$ $\mathcal{F}(\mathbf{S}^{-}(\mathbb{R}^m),\mathbf{S}^{-}(\mathbb{R}^m)$) decomposing  $\partial_i^v$ $f^n$ in $\mathbb{R}^m$, there exists a unique couple $(\mathbf{\beta}_1,\mathbf{\beta}_2)$ in $\mathbb{R}^{2m}$ such as: 
\begin{equation}
S_k(f) = \mathbf{\beta}_1 \Psi^{+}_k(f) + \mathbf{\beta}_2 \Psi^{-}_k(f), \qquad \forall k\in\mathbb{Z} 
\end{equation}
\vspace{1.0em}
\\ Two important results shown in \textit{\cite{JPMontillet2013} (Lemma and Theorem)} are: 
\vspace{1.0em}
\\$\bold{Lemma}$ $1$: For $f$ in $\mathbf{S}^{-}(\mathbb{R}^m)$, the family of DEO ${\Psi}_{k}^{+}$ ($k\in\mathbb{Z}$) decomposes $\partial_i^v$ $f^n$, $\forall v\in\mathbb{Z}^+ -\{ 0 \}$, $n\in\mathbb{Z}^+ -\{0,1\}$ and $i\in [1,...,m]$. 
\vspace{1.0em}
\\$\bold{Theorem}$ $1$: For $f$ in $\mathbf{s}^{-}(\mathbb{R}^m)$, the families of DEO ${\Psi}_{k}^{+}$ and ${\Psi}_{k}^{-}$ ($k\in\mathbb{Z}$) decompose uniquely $\partial_i^v$ $f^n$, $\forall v\in\mathbb{Z}^+-\{0\}$, $n\in\mathbb{Z}^+-\{0,1\}$ and $i\in [1,...,m]$. 
\vspace{1.0em}
\\ The $\bold{Lemma}$ $1$ and $\bold{Theorem}$ $1$  were then extended in \cite{JPMontillet2014} to the family of generalized operator with :
\vspace{1.0em}
\newline $\bold{Lemma}$ $2$: For $f$ in $\mathbf{S}_{p}^{-}(\mathbb{R}^m)$, $p$ in  $\mathbb{Z}^+$, the families of generalized energy operators $[[.]^{p}]_k^+$ ($k\in\mathbb{Z}$) decompose $\partial_i^v$ $([[f]^{p-1}]_1^+)^n$ $\forall v\in\mathbb{Z}^+-\{0\}$, $n\in\mathbb{Z}^+-\{0,1\}$ and $i\in [1,...,m]$. 
\vspace{1.0em}
\newline $\bold{Theorem}$ $2$: For $f$ in $\mathbf{s}_p^{-}(\mathbb{R}^m)$, for $p$ in $\mathbb{Z}^+$, the families of generalized operators $[[.]^p]_{k}^+$ and $[[.]^p]_{k}^-$ ($k\in\mathbb{Z}$) decompose uniquely $\partial_i^v$ $([[f]^{p-1}]_{1}^+)^n$ $\forall v\in\mathbb{Z}^+-\{0\}$, $n\in\mathbb{Z}^+-\{0,1\}$ and $i\in [1,...,m]$. 
\vspace{1.0em}
\\ $\mathbf{S}_{p}^{-}(\mathbb{R}^m)$ and $\mathbf{s}_{p}^{-}(\mathbb{R}^m)$ ($p$ in $\mathbb{Z}^+$) are energy spaces in $\mathbf{S}^-(\mathbb{R}^m)$ defined in the next section.
\vspace{0.5em}
\\ $\bold{Remark}$ $(5)$  One can extend the $\bold{Theorem}$ $1$, $\bold{Theorem}$ $2$, $\bold{Lemma}$ $1$ and $\bold{Lemma}$ $2$ for $f^n$ with $n$ in $\mathbb{Z}$ following previous discussions in \cite{JPMontillet2013} (Section 3, p.74) and \cite{JPMontillet2014} (Section $4$). $n$ is here restricted to $\mathbb{Z}^+ -\{0,1\}$ in order to easy the whole mathematical development. 
\subsection{Energy Spaces in $\mathbf{S}^{-}(\mathbb{R}^m)$} \label{energyspace}
Let us introduce the energy spaces and some properties.
\vspace{1.0em}
\\$\bold{Definition}$ $3$ \textit{\cite{JPMontillet2017}, Definition 3}: The energy space $\mathbf{E}_p \subsetneq \mathbf{S}^{-}(\mathbb{R}^m)$, with $p$ in $\mathbb{Z}^+$, is equal to $\mathbf{E}_p=\bigcup_{v\in\mathbb{Z}^+ \cup \{0\}} \mathbf{M}_p^v$. 
\vspace{1.0em}
\newline with $\mathbf{M}_p^v$ $\subsetneq \mathbf{S}^{-}(\mathbb{R}^m)$  for $v$ in $\mathbb{Z}^+$ defined as
\begin{equation}\label{energyspaceH}
\mathbf{M}_p^v=\{ g \in \mathbf{S}^{-}(\mathbb{R}^m) | \hspace{0.5em} g = \partial_i^k \big ( \big[ [f]^p \big ]_1^+ \big )^n , \big[ [f]^p \big ]_1^+ \in \mathbf{S}^{-}(\mathbb{R}^m),\hspace{0.5em} k \in \mathbb{Z}^+, \hspace{0.5em} \forall k \leq v ,  \hspace{0.5em} n\in\mathbb{Z}^+-\{0\}, \hspace{0.5em} i\in [1,...,m]\}
\end{equation}
%
%
%
\vspace{0.5em}
\\The energy spaces, $\mathbf{S}_p^{-}(\mathbb{R}^m)$ and $\mathbf{s}_p^{-}(\mathbb{R}^m)$ ($p$ $\in \mathbb{Z}^+$) , cited in $\bold{Lemma}$ $2$ and $\bold{Theorem}$ $2$  are defined:

\begin{eqnarray}
\mathbf{S}_p^{-}(\mathbb{R}^m)&=&\{ \mathbf{E}_p = \bigcup_{i\in\mathbb{Z}^+ \cup \{0\}} \mathbf{M}_p^i \} \nonumber \\
\mathbf{s}_p^{-}(\mathbb{R}^m) &=& \{ f \in  \mathbf{S}_p^{-}(\mathbb{R}^m) | f \notin  \cup_{i\in [1,...,m]} \big ( \cup_{k \in \mathbb{Z}} Ker([[f]^p]_{k,i}^+)\cup(\cup_{k \in \mathbb{Z}-\{1\}} Ker([[f]^p]_{k,i}^- \big)\} \nonumber \\
\end{eqnarray}
$\bold{Remark}$ $(6)$ $\bold{Definition}$ $3$ does not follow completely $\bold{Definition}$ $3$ in \cite{JPMontillet2017}, because the energy space  $\mathbf{M}_p^v$ is defined here $\forall k \leq v$, and only for $k=v$ in \cite{JPMontillet2017}.
%
\vspace{0.5em}
\\$\bold{Remark}$ $(7)$  In the previous definition, $\mathbf{M}_p^\infty = \{0\}$ ($\forall p \in \mathbb{Z}^+$). Also,  $\mathbf{M}_p^\infty \subset \mathbf{S}_p^{-}(\mathbb{R}^m)$, whereas $\mathbf{M}_p^\infty  \not\subset \mathbf{S}_p^{-}(\mathbb{R}^m)$   in \cite{JPMontillet2017} and \cite{JPMontillet2014}. The inclusion does not change $\bold{Lemma}$ $2$ and $\bold{Theorem}$ $2$ (i.e. $\mathbf{M}_p^\infty \not \subset \mathbf{s}_p^{-}(\mathbb{R}^m)$). The justification of not including this space was only based on the applications of the theory in  \cite{JPMontillet2017} and \cite{JPMontillet2014} which is not justified in this work.
\vspace{0.5em}
\\We can now state some properties associated with the energy spaces on $\mathbf{S}^{-}(\mathbb{R}^m)$. 
\vspace{0.5em}
\\$\textcolor{black}{\bold{Properties}}$ $2$: $\forall$ $v$ in  $\mathbb{Z}^+$, and in particular $v_1$, $v_2$ in  $\mathbb{Z}^+$ (with  $v_1 < v_2$), $p$ in $\mathbb{Z}^+$, we have the  following inclusions:
\begin{itemize}
\item[1-] $\mathbf{M}_p^{v} \subsetneq$ $\mathbf{H}^{v}(\mathbb{R}^m)$
\item[2-] $\mathbf{M}_p^{v_2} \subsetneq \mathbf{M}_p^{v_1}$
\item[3-] $\mathbf{E}_p =\bigcup_{v\in\mathbb{Z}^+ \cup \{0\}} \mathbf{M}_p^v \subsetneq  \mathbf{H}^{0}(\mathbb{R}^m)$
\end{itemize}
\begin{proof}
\circled{1} Let us recall the definition of the Hilbert space on $\mathbb{R}^m$ according to $\bold{Appendix}$ $I$ , $\mathbf{Definition}$ $I.1$ and $\mathbf{Definition}$ $1$. 
\begin{equation}
\mathbf{H}^{v}(\mathbb{R}^m) =\{f \in L^2(\mathbb{R}^m)\lvert D^{\alpha} f \in L^2(\mathbb{R}^m), \forall \abs{\alpha} \leq v \}
\end{equation}
Looking at the definition of the energy space $\mathbf{M}_p^{v}$ and $\mathbf{H}^{v}(\mathbb{R}^m)$, one can notice the similitude. However , the multi-index derivative $D^{\alpha}$ (\cite{srx})contains also the cross-derivatives (e.g., $\frac{\partial^2}{\partial t_1 \partial t_2}$), whereas there are no cross-derivatives in the definition of $\partial_i^v$ at the beginning of $\bold{Appendix}$ $I$. Thus, the energy spaces $\mathbf{M}_p^{v}$ ($p$ $\in$ $\mathbb{Z}^+$, $v$ $\in \mathbb{Z}^+-\{0\}$) is defined without the cross-derivatives. In addition with $\bold{\textcolor{black}{Properties}}$ $1$, $\mathbf{S}^{-}(\mathbb{R}^m) \subsetneq L^2(\mathbb{R}^m)$. 
 Thus, by definition we have the relationship $\mathbf{M}_p^{v} \subsetneq$ $\mathbf{H}^{v}(\mathbb{R}^m)$
\vspace{0.5em}
\\ \circled{2} With $\bold{Remark}$ $(3)$, we know that for $v_1 < v_2$, 
$\mathbf{H}^{v2}(\mathbb{R}^m) \subset \mathbf{H}^{v1}(\mathbb{R}^m)$. Now, with \circled{1} , $\mathbf{M}_p^{v_1} \subsetneq$ $\mathbf{H}^{v_1}(\mathbb{R}^m)$ and $\mathbf{M}_p^{v_2} \subsetneq$ $\mathbf{H}^{v_2}(\mathbb{R}^m)$. Now by definition of $\mathbf{M}_p^{v_1}$ and  $\mathbf{M}_p^{v_2}$,  $\mathbf{M}_p^{v_1} \bigcap \mathbf{H}^{v2}(\mathbb{R}^m)$ $=$ $\mathbf{M}_p^{v_2}$.
Finally, $\mathbf{M}_p^{v_2} \subsetneq \mathbf{M}_p^{v_1}$.
\vspace{0.5em}
\\ \circled{3} From $\bold{Remark}$ $(3)$, $\mathbf{H}^{0}(\mathbb{R}^m) = L^2(\mathbb{R}^m)$, $\mathbf{S}^{-}(\mathbb{R}^m) \subset L^2(\mathbb{R}^m)$ and (by definition of the energy space) $\mathbf{E}_p =\bigcup_{v\in\mathbb{Z}^+ \cup \{0\}} \mathbf{M}_p^v \subsetneq \mathbf{S}^{-}(\mathbb{R}^m)$ ($p \in \mathbb{Z}^+$). Thus, $\mathbf{E}_p =\bigcup_{v\in\mathbb{Z}^+ \cup \{0\}} \mathbf{M}_p^v \subsetneq  \mathbf{H}^{0}(\mathbb{R}^m)$ ($p \in \mathbb{Z}^+$).
\end{proof}
Furthermore, $\bold{Appendix}$ $III$ discusses the relationship between the subspaces $\mathbf{M}_p^{v}$ and $\mathbf{M}_{p-1}^v$ ($p$ $\in$ $\mathbb{Z}^+$).
\vspace{0.5em}
\\ Finally, because we are studying functions and operators in subspaces of $\mathbf{S}^{-}(\mathbb{R}^m)$ with  $\mathbf{S}^{-}(\mathbb{R}^m) \subset L^2(\mathbb{R}^m)$, one need to extend $\bold{Proposition}$ $1$ in \cite{JPMontillet2017} and \cite{JPMontillet2014}.  
\vspace{0.5em}
\newline \textit{$\bold{Proposition}$ $1$: If for $n$ $\in$ $\mathbb{Z}^+$, $f^n$ $\in$ $\mathbf{S}^{-}(\mathbb{R}^m)$ and analytic;  for any ($p_i$,$q_i$) $\in$ $\mathbb{R}^2$ and $\tau_i$  $ \in [q_i ,p_i]$ ($\forall i \in [1,...,m]$), and $\mathcal{E}(f^n)$ is analytic, where}
\begin{equation}\label{EnergyfunDefine02}
\mathcal{E}(f^n(\tau_i)) = \int_{q_i}^{\tau_i} \abs{f^n(t_i)}^2dt_i < \infty
\end{equation}
\textit{then} 
\begin{eqnarray}\label{refeq1}
\mathcal{E}(f^n(p_i)) &=& \mathcal{E}(f^n(q_i)) + \sum_{k=0}^\infty \partial_{t_i}^k (f^n(q_i))^2 \frac{(p_i-q_i)^k}{k!} <\infty \nonumber \\
\end{eqnarray}
\textit{is a convergent series.}
\begin{proof}
The proof of $\bold{Proposition}$ $1$ for $i$ equal $1$ is given in \cite{JPMontillet2017} (p.4). The extension of the proof for the case $i$ equal $m$ is straightforward with the general definition for any ($p_i$,$q_i$) $\in$ $\mathbb{R}^2$ and $\tau_i$  $ \in [q_i ,p_i]$ ($\forall i \in [1,...,m]$).  
\end{proof}
\section{Multiplicity of the Solutions in $\mathbf{S}^{-}(\mathbb{R}^m)$}\label{Multiplicity}
To recall \cite{JPMontillet2017}, a possible application of the theory of the energy operators is to look at solutions of a given partial differential equation for solutions in $\mathbf{S}^{-}(\mathbb{R}^m)$ of the form $\partial_i^v(f^n)$.  Instead of solving the equation for specific values (e.g., boundary conditions), the work in  \cite{JPMontillet2017} (  \cite{JPMontillet2017}, ${Theorem}$ $1$ and $corollary$)defines the concept of \textit{multiplicity} of solutions in $\mathbf{S}^{-}(\mathbb{R}^m)$ ($m\in[1,2]$) such as the study of the multiple solutions of a PDE based on the definition of the energy spaces $\mathbf{E}_p$ ($p \in \mathbb{Z}^+$).  One way to understand this concept, is to study the convergence of the development in Taylor series of the energy function associated to a nominated energy space. It was shown in \cite{JPMontillet2017} that taking into account additional terms of the Taylor series leads to define additional solutions of the wave equation (see  Section $4$ \cite{JPMontillet2017}). In this section, we extend this concept to $\mathbf{S}^{-}(\mathbb{R}^m)$ ($m \in \mathbb{N}_0$) and we reformulate the results from\cite{JPMontillet2017} for the solutions in the subspaces $\mathbf{M}_p^{v} \subset \mathbf{E}_p \subset \mathbf{S}^{-}(\mathbb{R}^m) \subset L^2(\mathbb{R}^m) $ ($p \in \mathbb{Z}^+$, $v \in \mathbb{Z}^+$).
\vspace{0.5em}
\\ \indent Let us define any PDEs of the form:
\begin{equation}\label{eqpdep}
 \left\{
\begin{array}{rl}
\sum_{j \in \mathbb{Z}^+} \sum_{i\in [0,..,m]} a_{ij} & \partial_i^{v_j} g  =0, \\
\forall g \in \mathbf{A}(\mathbb{R}^m) \subseteq & \mathbf{S}^{-}(\mathbb{R}^m), \\
\forall  a_{ij} \in \mathbb{R}, & \hspace{0.5em} v_j \in  \mathbb{Z}^+
%
\end{array} \right.
\end{equation} 
Thus, all the solutions are here defined in  $\mathbf{A}(\mathbb{R}^m) \subseteq  \mathbf{S}^{-}(\mathbb{R}^m)$. Now, we are interested in the solutions which can be defined on the energy spaces $\mathbf{E}_p$ ($p \in \mathbb{Z}^+$). In other words, $\mathbf{A}(\mathbb{R}^m) \bigcap_{p\in \mathbb{Z}^+}\mathbf{E}_p \neq \{\emptyset\}$. In particular, we choose the  solution $g=0 \in  \mathbf{A}(\mathbb{R}^m) \bigcap_{p\in \mathbb{Z}^+}\mathbf{E}_p$. Furthermore, one can define $g$ $\in \mathbf{A}(\mathbb{R}^m) \bigcap_{p\in \mathbb{Z}^+}\mathbf{E}_p$, such as $\exists v \in \mathbb{Z}^+$ for $g$ $\in \mathbf{A}(\mathbb{R}^m) \bigcap_{p\in \mathbb{Z}^+}\mathbf{M}^v_p$. In other words, $\exists f \in  \mathbf{S}^{-}(\mathbb{R}^m)$ and $n \in \mathbb{Z}^+-\{0\}$, such as $g= ([[f]^p]_1^+)^n$. 
\\ Now, one can then state a general theorem of \textit{multiplicty of solutions} based on  \cite{JPMontillet2017}. It follows:
\vspace{1em}
\newline $\bold{Theorem}$ $3$ (\textit{Multiplicity of Solutions in $\mathbb{R}^m$}):
If  $\mathbf{A}(\mathbb{R}^m) \subseteq  \mathbf{S}^{-}(\mathbb{R}^m)$ is a subspace of all the solutions of a nominated linear PDE. For $p \in \mathbb{Z}^+$,  $g$ is in  $\mathbf{E}_p$. Then, $g$ is a solution for this linear PDE if and only if:
\begin{enumerate}
\item [1.](General condition to be a solution)  $\mathbf{A}(\mathbb{R}^m) \bigcap \mathbf{E}_p \neq \{\emptyset\}$.
%
%
\item [2.] (Solutions in $\mathbf{S}^-(\mathbb{R}^m)$ ) $g \in$ 
$\mathbf{A}(\mathbb{R}^m) \bigcap \mathbf{E}_p$, 
$ \exists m \in \mathbb{R}^+$ such as  $m =$ $sup$ $(\mathcal{E} (g) )$. 
\item [3.] (Multiplicity of the solutions)
If $g$ $\in \mathbf{M}^v_p$ ($v \in \mathbb{Z}^+$), $\exists f \in  \mathbf{S}^{-}(\mathbb{R}^m)$ and $n \in \mathbb{Z}^+-\{0\}$, such as $g= \partial_i^v ([[f]^p]_1^+)^n$ ($i \in [0,..,m]$)and $\forall k \geq v$, $k \in \mathbb{Z}^+$,  $\partial_i^k ([[f]^p]_1^+)^n \in \mathbf{A}(\mathbb{R}^m) \bigcap \mathbf{E}_p$. 
\item [4.](Superposition of solutions and energy conservation ) If $F  \in \mathbf{A}(\mathbb{R}^m) \bigcap \mathbf{E}_p$, with $F = \sum_{ k\in \mathbb{Z}^+,\forall k \geq v } \partial_i^k([[f]^p]_1^+)^n$ such as $\partial_i^k([[f]^p]_1^+)^n \in \mathbf{M}_p^k$ ($i \in [0,..,m]$), then $\mathcal{E}(F) < \infty$.
\end{enumerate}
\begin{proof}
The proof is the generalization of what was already written in \cite{JPMontillet2017} (see $Theorem$ $2$ in \cite{JPMontillet2017}) for the case $m$ equal $1$. Here is the generalization to $m$.
\begin{enumerate}
\item [1.] This is the definition of a solution for a nominated PDE with solutions in $\mathbf{A}(\mathbb{R}^m)$  and in the energy space $\mathbf{E}_p$.
\item [2.]  $g \in$ 
$\mathbf{A}(\mathbb{R}^m) \bigcap \mathbf{E}_p \subset L^2(\mathbb{R}^m)$, thus $\mathcal{E}(g) < \infty$. With $\bold{Proposition}$ $1$, it means that for any ($p_i$,$q_i$) $\in$ $\mathbb{R}^2$ and $\tau_i$  $ \in [q_i ,p_i]$ ($\forall i \in [1,...,m]$)
\begin{equation}\label{EnergyfunDefine02io}
\mathcal{E}(g(\tau_i)) = \int_{q_i}^{\tau_i} \abs{g(t_i)}^2dt_i < \infty
\end{equation} 
Thus, following \cite{UCDavis}, one can define $m_i$ $\in$ $\mathbb{R}$ such as $m_i = sup_{\tau_i \in [q_i ,p_i]} \mathcal{E}(g(\tau_i))$
and then we define $m=max_{i \in [1,...,m]} m_i$. With our notation, it is equivalent to write  $m =$ $sup$ $(\mathcal{E} (g) )$.
\item [3.] It is sufficient to show that for $v \in \mathbb{Z}^+$, $\forall k \geq v$, $\mathbf{A}(\mathbb{R}^m) \bigcap \mathbf{M}_p^k \neq \{\emptyset\}$.
\\ Now, with the definition  $\mathbf{A}(\mathbb{R}^m) \bigcap\mathbf{E}_p \neq \{\emptyset\}$, and  $\mathbf{A}(\mathbb{R}^m) \bigcap \mathbf{M}_p^v \neq \{\emptyset\}$. In addition, $\mathbf{M}_p^\infty =\{0\}$, $\mathbf{M}_p^\infty \subset \mathbf{M}_p^k$ ($\forall k \geq v$) and $0 \in \mathbf{A}(\mathbb{R}^m) \bigcap\mathbf{M}_p^k $. 
The interest of this statement is the function $\partial_i^v h\in \mathbf{S}^-(\mathbb{R}^m)$ such as $\exists$ $k \in$ $\mathbb{Z}^+$ with $k\geq v$ and $\partial^{k}_i h =0$. In particular, if we introduce a numerical approximation in order to get the condition $\partial^k_i h  \sim 0$. In other words, 
\begin{equation}\label{approximate01}
\partial^{k}_i h  \sim 0 \leftrightarrow \{ \exists \hspace{0.5em} \epsilon \in \mathbb{R}^+, \hspace{0.5em} \epsilon <<1, \hspace{0.5em} \forall k \in \mathbb{Z}^+, \hspace{0.5em} k>0, \hspace{0.5em} such \hspace{0.5em} as  \hspace{0.5em}    |\partial^{k}_i h | \leq \epsilon \}
\end{equation}
In some examples in Section $4$ in \cite{JPMontillet2017} and Section $6$ in  \cite{JPMontillet2014}, it is shown that the evanescent waves when solving the wave equation for specific solutions, is a particular example of those functions.
\item [4.] The proof follows \cite{JPMontillet2017}($Theorem$ $2$). This statement is to guarantee that there is a finite sum of energy with the superposition of multiple solutions.  Thus with the development in statement (2.), one can use the Minkowski inequality (e.g,  \cite{Hardy}, $Theorem$ $202$) for $\tau_i$ in $[p_i,q_i]$ ($\forall i \in [1,...,m]$)
\begin{eqnarray}
\mathcal{E} (F(\tau_i)) &=& \int_{p_i}^{\tau_i} |\sum_{ k\in \mathbb{Z}^+,\forall k \geq v} \partial_i^k ([[f(t_i)]^p]_1^+)^n|^2 dt_i \nonumber \\
\big ( \mathcal{E} (F(\tau_i)) \big )^{0.5} & \leq & \sum_{ k\in \mathbb{Z}^+,\forall k \geq v} \big (\int_{p_i}^{\tau_i} |\partial_i^k ([[f(t_i)]^p]_1^+)^n|^2 dt_i \big )^{0.5} \nonumber \\
\big ( \mathcal{E} (F(\tau_i)) \big )^{0.5} & \leq & \sum_{ k\in \mathbb{Z}^+,\forall k \geq v} \big ( \mathcal{E} (\partial_i^k ([[f(\tau_i)]^p]_1^+)^n) \big )^{0.5}\nonumber \\
\big ( \mathcal{E} (F(\tau_i)) \big )^{0.5} & \leq & \sum_{ k\in \mathbb{Z}^+,\forall k \geq v} m_k^{0.5}
\end{eqnarray} 
with $ m_k$ $=sup_{\tau_i \in [q_i ,p_i]}$ $( \mathcal{E} (\partial_i^k ([[f(\tau_i)]^p]_1^+)^n) )$.
Thus, (4.) stands if and only if $\sum_{ k\in \mathbb{Z}^+,\forall k \geq v} m_k^{0.5} < \infty$. As $\forall$ $k\in \mathbb{Z}^+$, $\forall k \geq v$, $m_k$ is in $\mathbb{R}^+$, it then exists $M =sup \sum_{k\in \mathbb{Z}^+,\forall k \geq v} m_k^{0.5}$. One possibility is  $ \exists k_o$ in $\mathbb{Z}^+$ such as $\forall k > k_o$, then $m_k =0$.
\end{enumerate}

\end{proof}
\section{Some Applications}\label{Applications}
This section focuses on the application of the energy space theory. The first section is the study of the concept of \textit{multiplicity of solutions} with a simple mathematical example using the wave equation. Then, the second section is discussing the application of this concept within the Woodward effect \cite{Woodward, Woodwardb}
\subsection{Energy variation and wave equation}
As a simple case of linear PDE, the wave equation with the particular solutions of the form of evanescent waves, was already discussed in  Section $6$ of \cite{JPMontillet2014} and \cite{JPMontillet2017}. However, it is an interesting example to apply and understand the concept of multiplicity  stated in $\bold{Theorem}$ $3$. From \cite{Petit} or \cite{Amzallag}, the wave equation can be formulated in $\mathbb{R}^2$ (with $t$ and $r$ the time and space variables):
\begin{equation}\label{eqpdep}
 \left\{
\begin{array}{rl}
\partial_r^{2}g(r,t) &  - \frac{1}{c^2} \partial_t^{2}g(r,t) = 0 ,\\
or \hspace{0.5em} \square g(r,t) = 0, \\
t \in [0, T ], \hspace{0.5em} r \in [r_1, r_2], & \hspace{0.5em} (r_1,r_2, T) \in \mathbb{R}^3, \hspace{0.5em} r_1<r_2\\
 t_0 \in [0, T ], \hspace{0.5em} r_0 \in [r_1, r_2] \\
\end{array} \right.
\end{equation}
$c$ is the speed of light. Note that the values of $t$ and $r$ are restricted to some interval, because it is conventional to solve the equation for a restricted time interval in $\mathbb{R}^+$ and a specific region in space. 
According to the previous section, we are here interested in the solutions in the energy (sub)space $\mathbf{M}_p^k$, of the kind $g(r,t) = \partial_t^k ([[f]^p]_1^+)^n(r,t)$ ($n$ in $\mathbb{Z}^+-\{0\}$, $p$ in $\mathbb{Z}^+$, $k$ in $\mathbb{Z}^+$). Furthermore, the relationship $\mathbf{M}_p^k \subset \mathbf{S}^{-}(\mathbb{R}^2) \subset L(\mathbb{R}^2)$ imposes that the solutions should be finite energy functions, decaying for large values of $r$ and $t$. It was previously underlined in \cite{JPMontillet2014} and \cite{JPMontillet2017} that planar waves should be rejected, because this type of solutions does not belong to $L(\mathbb{R}^2)$. However, evanescent waves are a type of solutions included in $\mathbf{S}^{-}(\mathbb{R}^2)$ and   considered in this work. They are here defined such as:
\begin{equation}\label{evanescent}
 \left\{
\begin{array}{rl}
f(r,t) = & Real \{ A \exp{(u_2 r)} \exp{(\textcolor{black}{i}(\omega t -u_1 r))} \} ,\\
t \in [0, T ], \hspace{0.5em} r \in [r_1, r_2], & \hspace{0.5em} (r_1,r_2) \in \mathbb{R}^2, \hspace{0.5em} r_1<r_2\\
\end{array} \right.
\end{equation}
\textcolor{black}{$i^2=-1$}, $u_1$ and $u_2$ are the wave numbers,  $\omega$ is the angular frequency and $A$ is the amplitude of this wave \cite{Petit}. Assuming $\omega$ and ($u_1$, $u_2$) known, one can add some boundary conditions in order to estimate $u_1$, $u_2$ and $A$. Furthermore, a traveling wave solution of \eqref{eqpdep} should satisfy the dispersion relationship  between $u_1$, $u_2$ and $\omega$  \cite{Petit}. However, our interest is just the general form assuming that all the parameters are known. For $p=0$, the type of solutions in $\mathbf{M}_0^k$ are :
\textcolor{black}{
\begin{equation}\label{evanescentS1}
 \left\{
\begin{array}{rl}
 \partial_t^k f^n (r_0,t) = & (i\omega n)^k  f^n(r_0,t)\} ,\\
 \partial_r^k f^n (r,t_0) = & (n(u_2-i u_1))^k f^n(r,t_0)\},\\
t \in [0, T ], \hspace{0.5em} r \in [r_1, r_2], & \hspace{0.5em} (r_1,r_2, T) \in \mathbb{R}^3, \hspace{0.5em} r_1<r_2\\
 t_0 \in [0, T ], \hspace{0.5em} r_0 \in [r_1, r_2], \hspace{0.5em} n \in \mathbb{Z}^+ -\{0\}, \hspace{0.5em} k \in \mathbb{Z}^+-\{0\} \\
\end{array} \right.
\end{equation}
}
In $\mathbf{M}_1^k$, one can then write the type of solutions
\textcolor{black}{
\begin{eqnarray}\label{evanescentS2Psi}
\partial_t^k \Psi_{1,t}^{+}(f) (r_0,t) &=&\partial_t^k ([[f(r_o,t)]^1]_{1,t}^{+}\nonumber \\
\partial_t^k \Psi_{1,t}^{+}(f) (r_0,t) &=& Real \{ (i 2 k \omega ) f^2 (r_0,t)\} \nonumber \\
\partial_r^k \Psi_{1,r}^{+}(f)(r,t_0) &=& Real \{  (2k(u_2-iu_1)) f^2 (r,t_0)\} \nonumber \\
t \in [0, T ], && \hspace{0.5em} r \in [r_1, r_2],  \hspace{0.5em} (r_1,r_2, T) \in \mathbb{R}^3, \hspace{0.5em} r_1<r_2 \nonumber \\
t_0 \in [0, T ], && \hspace{0.5em} r_0 \in [r_1, r_2], \hspace{0.5em} n \in \mathbb{Z}^+ -\{0\}, \hspace{0.5em} k \in \mathbb{Z}^+-\{0\}
\end{eqnarray}
}
Let us consider the form of solutions which propagates in a closed cavity (e.g., closed wave guide \cite{Petit}). One possible solution is the evanescent wave described in \eqref{evanescent}. Now, if $f$ and $\mathcal{E}(f)$ are analytic in $\mathbb{R}^2$, with $\bold{Proposition}$ $1$   we can  assume that $f$ is finite energy (and more generally in $\mathbf{S}^-(\mathbb{R}^2)$) with a wise choice on the parameters $A$, $u_1$, $u_2$ and $\omega$. One can estimate the difference of energy in time over $dt$ inside  the cavity at a specific location $r_0$ ($r_0$ in $[r_1, r_2]$) such as
\begin{eqnarray}\label{refeq100}
\mathcal{E}(f(r_0,T)) & =& \int_0^{T} (f(r_0,h))^2dh < \infty \nonumber \\
\mathcal{E}(f(r_0,T+dt)) &=& \mathcal{E}(f(r_0,T)) + \sum_{k=0}^\infty \partial_t^k (f^2(r_0,T)) \frac{(dt)^k}{k!} <\infty \nonumber \\
%
%
\mathcal{E}(f(r_0,T+dt)) &=& \mathcal{E}(f(r_0,T)) + f^2(r_0,T)dt + \sum_{k=1}^\infty \partial_t^{k-1}  \Psi_{1,t}^{+}(f)(r_0,T) \nonumber \\
%
\mathcal{E}(f(r_0,T+dt)) & \simeq & \mathcal{E}(f(r_0,T)) + f^2(r_0,T)dt 
\end{eqnarray}
Here the symbol '$\simeq$' means that 
%
\begin{equation}\label{approximate01}
\exists \hspace{0.5em} \epsilon \in \mathbb{R}^+, \hspace{0.5em} \epsilon <<1, \hspace{0.5em} \forall k \in \mathbb{Z}^+, \hspace{0.5em} k>0 | \hspace{0.5em}    |\partial_t^{k-1} \big ( \Psi_{1,t}^{+}(f)(r_0,T)  \big ) | < \epsilon |f^2(r_0,T)|
\end{equation}
Now, let us do a hypothesis that $\mathcal{E}(f(r_0,T+dt))$ increases significantly over $dt$ modifying the approximation in \eqref{approximate01}
\begin{equation}\label{approximate01p}
\exists \hspace{0.5em} \epsilon \in \mathbb{R}^+, \hspace{0.5em} \epsilon <<1, \hspace{0.5em} \forall k \in \mathbb{Z}^+, \hspace{0.5em} k>1 | \hspace{0.5em}    |\partial_t^{k-1}  \Psi_{1,t}^{+}(f)(r_0,T) | < \epsilon | \Psi_{1,t}^{+}(f)(r_0,T)|
\end{equation}
and then,
 \begin{eqnarray}
 \mathcal{E}(f(r_0,T+dt)) & \simeq & \mathcal{E}(f(r_0,T)) + f^2(r_0,T)dt +  \Psi_{1,t}^{+}(f)(r_0,T) \frac{dt^2}{2} \nonumber \\
 \end{eqnarray}
To recall that $f^2(r,t) \in \mathbf{M}_0^0$, $ \partial_t f^2(r,t) \in \mathbf{M}_0^1$ and $\Psi_{1,t}^{+}(f)(r,t)  \in \mathbf{M}_1^0$, and using $\bold{Theorem}$ $3$, one can take into account solutions in those subspaces.
The multiplicity of the solutions due to the variation of energy  can be formulated as an approximation for taking into account additional solutions produced by the wave equation. 
\vspace{0.5em}
\\ $\bold{Remark}$ ($8$): In \cite{JPMontillet2017}, the general idea was to look for the solutions of linear PDEs in $\mathbf{S}^-(\mathbb{R})$ associated with energy subspaces $\mathbf{s}_p^-(\mathbb{R})$ ($p \in$ $\mathbb{Z}^+$) in order to apply $\bold{Theorem}$ $1$ in \cite{JPMontillet2017}, which is here generalized in $\bold{Theorem}$ $3$ for $\mathbb{R}^m$ ($m \in$ $\mathbb{N}^+$). The purpose was to find the subspaces reduced to $\{0\}$ when studying the convergence of the Taylor series of the energy functions. However, the redefinition of the energy subspaces $\mathbf{M}_p^k$ within the Sobolev spaces defined in Section \ref{energyspace} allows us to look for solutions in $\mathbf{S}_p^-(\mathbb{R}^m)$ in order to use $\bold{Lemma}$ $2$. Because of the inclusion of the energy spaces shown in $\bold{Properties}$ $2$ using the Sobolev embedding (e.g.,  $\bold{Theorem}$ $I.1$ in $\bold{Appendix}$ $I$) such as  $ \mathbf{M}_p^{k+1}k\subset \mathbf{M}_p^k $ ($k \in \mathbb{Z}^+$, $p\in \mathbb{Z}^+$), $\mathbf{M}_p^\infty =\{0\} \subset \mathbf{M}_p^k$. 
\vspace{0.5em}
\\ $\mathbf{Definition}$ $4$ \textit{(Energy Parallax)}  Considering a linear PDE with some solutions in $\mathbf{A}(\mathbb{R}^m)$ such as $\mathbf{A}(\mathbb{R}^m) \bigcap \mathbf{S}^-(\mathbb{R}^m) \neq \{\emptyset\}$. Furthermore, if it exists $p$ and $v$ $\in$ $\mathbb{Z}^+$ such as  $\mathbf{A}(\mathbb{R}^m) \bigcap \mathbf{M}_p^v  \neq \{\emptyset\}$, then we associate the energy $\mathcal{E}(f)$ for $f$ $\in$ $\mathbf{A}(\mathbb{R}^m) \bigcap \mathbf{M}_p^v$, such as one can estimate the variation $d\mathcal{E}(f)$ $=$ $(\mathcal{E}(f(q+dq))-\mathcal{E}(f(q)))$$dq$ over an elementary quantity $dq$ (e.g., space or time). 
If $d\mathcal{E}(f)$ is not negligible ($\exists \epsilon$ $\in \mathbb{R}^m$ such as $\epsilon >>1$ and $d\mathcal{E}(f) > \epsilon$ ), then one can consider additional solutions in $\mathbf{A}(\mathbb{R}^m) \bigcap \mathbf{M}_p^{v+1}$.
\subsection{Variation of EM energy density and the Woodward effect}\label{sectionEMvariation}
In this section, the theory of energy space is applied to the possible  variations of electromagnetic energy density due to, for example, skin depth effect \cite{Amzallag} inside some conductive material. Beyond this application, the interest is to give a physical meaning of taking into account those additional solutions in various energy spaces. The second part is dedicated to the Woodward effect and the possible relationship with the variation of EM energy density in some specific settings.
\subsubsection{Variation of EM energy density}
Thus, let us formulate the variation in time of energy density ($u$)at the second order with a Taylor series development such as:
\begin{equation}\label{dweuqtion}
d u = \partial_t u \hspace{0.2em} dt + \partial_t^2 u \hspace{0.2em} \frac{dt^2}{2} + o(dt^2)
\end{equation}
$o$ is the Landau notation to omit higher order quantities. Note that at the first order $\frac{d u}{dt} =\partial_t u$.
The higher orders term are based on the assumptions that the EM waves inside the skin layer of the copper plate are evanescent waves and thus functions in the Schwartz space ($\mathbf{S}^-(\mathbb{R}^4)$ - with $3$ dimension variables and considering also the time ) \cite{Petit}. As discussed before, those solutions are finite energy functions and in $L(\mathbb{R}^4)$ (i.e.following \cite{JPMontillet2014} and \cite{JPMontillet2017},  $u=$ $\mathcal{E}(f(xo,yo,zo,T)) <\infty$ at some given point in the skin layer defined by the coordinates $xo,yo,zo$). 
Now, using the $\bold{Lemma}$ $1$  and the space $\mathbf{M}_0^k$ in Section \ref{energyspace}, we can state in  $\mathbf{S}^-(\mathbb{R}^4)$ 
\begin{eqnarray}\label{energyspaceH0022}
\mathbf{M}_0^k&=&\{ g \in \mathbf{S}^{-}(\mathbb{R}^4) | \hspace{0.5em} g = \partial_t^k  f^n(x_0,y_0,z_0,t ) \nonumber \\
&=& \alpha_n (\partial_t^{k-1} f^{n-2}(x_0, y_0,z_0,t ) (\Psi_1^+(f(x_0,y_0,z_0,t )))  \nonumber \\
, f \in \mathbf{S}^{-}(\mathbb{R}^4), \hspace{0.5em} n\in\mathbb{Z}^+-\{0\}, &&\hspace{0.5em} \alpha_n \in \mathbb{R}, \hspace{0.5em}  z_0 \in [0,L], \hspace{0.5em} (x_0,y_0) \in [0,a]^2 \}
\end{eqnarray}
Here $f$ is either the electric or magnetic field (i.e. the absolute norm of $\vec{E}$ and $\vec{B}$ respectively). With the concept of \textit{multiplicity of solutions} (e.g., $\bold{Theorem}$ $3$). If $g$ is a general solution of some linear PDEs, then $f^n$ can be identified as a special form of the solution (conditionally to its existence ). 
\\ Now considering the wave equation,  the electric field and magnetic field  are solutions and belong to the subspace $\mathbf{M}_0^k$ and associated with the variation of energy density $\partial_t u$. Furthermore, we can consider  the solutions in $\mathbf{M}_0^1$ associated with the variation of energy density $\partial_t^2 u$, which can be explained with the concept of $\it{multiplicity}$ of solutions. The solutions of interest in $\mathbf{M}_0^1$ are for the electric field $g =\partial_t E$ and the magnetic field $g =  \partial_t {B}$. 
%
The Taylor Series development of the energy of (for example) the electric field on a nominated position in space (i.e., $x_0,y_0,z_0$) and in an increment of time $dt$:
\begin{eqnarray}\label{refeq100}
%
\mathcal{E}(E(x_0,y_0,z_0,T+dt)) &=& \mathcal{E}(E(x_0,y_0,z_0,T)) + \sum_{k=0}^\infty \partial_t^k (E^2(x_0,y_0,z_0,T)) \frac{(dt)^k}{k!} <\infty \nonumber \\
d \mathcal{E}(E(x_0,y_0,z_0,T+dt)) &=&   \sum_{k=0}^\infty \partial_t^k (E^2(x_0,y_0,z_0,T)) \frac{(dt)^k}{k!}  \nonumber \\
%
%
\end{eqnarray}
Finally one can write the relationship with the energy density following \eqref{dweuqtion} and the previous Taylor series development for the electric and magnetic field:
\begin{eqnarray}
0.5 \big (\epsilon_0 \hspace{0.3em} \frac{d \mathcal{E}(E(x_0,y_0,z_0,T+dt))}{dt} +\frac{1}{\mu_0} \hspace{0.3em} \frac{d \mathcal{E}(B(x_0,y_0,z_0,T+dt))}{dt} \big ) &=& \nonumber \\
 && \hspace{-6em} 0.5 \big (\epsilon_0 E^2(x_0,y_0,z_0,T) \nonumber \\
+\frac{1}{\mu_0} B^2(x_0,y_0,z_0,T)) + \partial_t u \hspace{0.3em} \frac{dt}{2} + \partial_t^2 u \hspace{0.3em} \frac{dt^2}{6} +o(dt^2)&&
\end{eqnarray}
Therefore, taking into account the second order term of the energy density $\partial_t^2 u$ means that additional solutions should also be considered in the EM modeling. Note that in $\bold{Appendix}$ $IV$, we are taking an example of evanescent waves inside a copper wall (i.e. skin depth effect \cite{Amzallag}) and try to give further meaning to the consideration of higher order derivatives of the EM energy density where the additional solutions are defined with the energy spaces (e.g.,   $\partial_t {E} $ and  $\partial_t {B} $ in $\bold{M}_0^1$)
\subsubsection{ Derivation of the Woodward effect Using the electromagnetic energy density}\label{Woodwardsection}
%
%
This section focuses on the derivation of the Woodward effect created in a asymmetric EM cavity (i.e. frustum) due to EM waves reflected on the cavity's wall. Thus, the assumption is that the EM energy density variation results from the evanescent waves taking place in the skin depth of the asymmetric EM cavity's walls.
\subsubsection*{ Assumptions with the energy momentum relationship}
When the Woodward effect was established in \cite{Woodward, Woodwardb}, the authors implicitly assumed the rest mass of the piezoelectric material via the famous Einstein's relation in special relativity $\mathcal{E}= m c^2$ ($\mathcal{E}$ the rest energy associated with the rest mass $m$) and its variation via electrostrictive effect.

Here, the system is the asymmetric EM cavity. The rest mass is all the particles within it at time $t_0$ when no charges are on the cavity's walls. It excludes the photons considered with a null mass. Thus, the main assumption is that the EM excitation on the walls creates electric charges (i.e. electrons) which makes the rest mass varying with time. This assumption is the same as the mass variation of a capacitor between the charge and discharge times \cite{Porcelli}. It allows us to state the variation of rest energy such as:
\begin{eqnarray}
d \mathcal{E} &=& \mathcal{E}(t+dt) -\mathcal{E}(t) \nonumber \\
         &=& (m(t+dt) -m(t)) c^2  \nonumber \\
         &=&  d m c^2
\end{eqnarray}
Finally, the variation of rest energy $d \mathcal{E}$ is assumed to be equal to the variation of EM energy density ($d u$) resulting from the charges within the skin depth of the  walls. We neglect any electrostrictive effects compared to the variation of EM energy density. 

Note that at the particle level, the rest mass should satisfy the energy momentum relationship ($u_e$) for a free body in special relativity \cite{Moeller}:  
\begin{eqnarray}
u_e^2 &=& (pc)^2 +(m_e c^2)^2 \nonumber \\
 p &=& v \frac{u_e}{c^2} 
\end{eqnarray}
with $p$ the momentum  and $m_e$ the rest mass of the particle associated with the total energy $u_e$. The particle is accelerated via the Lorentz force applied to the whole cavity with obviously $v<<c$. Thus, we have also the relationship $ p^2 < (u_e/c)^2$.
\vspace{0.5em}
\\In the remainder, we also use the elementary variation $\delta$ which becomes $d$ for an infinitesimally small variation.
%
\subsubsection*{Derivation of the Woodward effect and relationship with EM energy density }

%
%
If we define the mass density such as $\rho = m/V$, then from \cite{Woodwardb}, one can write the elementary mass variation per unit of volume
\begin{eqnarray}
 \delta \rho  &=& \frac{\delta m}{V} \nonumber \\
 &\sim & d \rho \hspace{0.5em}(infinitesimally\hspace{0.5em} small\hspace{0.5em} variation)\nonumber \\
 d\rho  &=&  \frac{1}{4 \pi G}\big [ \frac{1}{\rho} \partial_t^2 \rho -\frac{1}{\rho^2} (\partial_t \rho)^2\big]\nonumber \\
\end{eqnarray}
Let us define the the rest energy $\mathcal{E}=\rho c^2$, then
\begin{eqnarray}
 d \rho  &=&  \frac{1}{4 \pi G}\big [ \frac{1}{\rho c^2} \partial_t^2 \mathcal{E} -\frac{1}{(\rho c^2)^2} (\partial_t \mathcal{E})^2\big]\nonumber \\
 d \rho  &=&  \frac{1}{4 \pi G}\big [ \frac{1}{\mathcal{E}} \partial_t^2 \mathcal{E}-\frac{1}{(\mathcal{E})^2} (\partial_t \mathcal{E})^2\big]\nonumber \\
\end{eqnarray} 
Now in some particular cases such as an EM cavity,   we assume that  the variation in time of the rest energy is equal to the variation of EM energy density $u$ (i.e. $\partial_t \mathcal{E} \simeq \partial_t u$ ), but the rest energy is much bigger than the EM energy density $\mathcal{E} >> u$. It allows then to state the relationship between the Woodward effect and the EM energy density
\begin{equation}\label{EMcoupling}
d \rho  =  \frac{1}{4 \pi G}\big [ \frac{1}{\mathcal{E}} \partial_t^2 u -\frac{1}{(\mathcal{E})^2} (\partial_t u)^2\big]
\end{equation}
The EM energy density $u$ follows the general definition of the sum of energy density from the electric ($u_E$) and magnetic ($u_B$)fields \cite{Petit}. Finally, \eqref{EMcoupling} can be seen as the definition of the EMG.
\section{Conclusions}
This work generalizes in the Schwartz space $\mathbf{S}^-(\mathbb{R}^m)$, the framework on conjugate Teager-Kaiser energy operators established in \cite{JPMontillet2013} and \cite{JPMontillet2014} for the case $m$ in $[1,2]$. The concept of \textit{multiplicity of solutions} defined in \cite{JPMontillet2017} is also redefined here in $\mathbf{Theorem}$ $3$. However, this concept uses the notion of energy spaces (${\mathbf{M}_p}^v$ ($p$ $\in \mathbb{Z}^+$, $v$ $\in \mathbb{Z}^+$), subspaces of $\mathbf{S}^-(\mathbb{R})$ \cite{JPMontillet2017}, \cite{JPMontillet2014} . In order to generalize their definition as subspaces of $\mathbf{S}^-(\mathbb{R}^m)$, the theory has been extended to some properties on the Hilbert spaces ($\mathbf{H}^v_1(\mathbb{R}^m)$) on $L^2(\mathbb{R}^m)$. In particular, we show in $\mathbf{Property}$ $2$ that ${\mathbf{M}_p}^v_2 \subsetneq {\mathbf{M}_p}^v_1$ ($v_1 <v_2$) and the inclusion ${\mathbf{M}_p}^v_1 \subsetneq \mathbf{H}^v_1(\mathbb{R}^m)\subsetneq L^2(\mathbb{R}^m)$.
\\ The concept of \textit{multiplicity of solutions} focuses on, generally speaking,  looking for solutions of a given linear PDE specifically in the energy spaces. In this way, it is not following the classical way of solving a linear PDE with boundary conditions. Three examples illustrate this concept. The first one  investigates some type of solutions (e.g., evanescent waves) of the wave equation when analysing the Taylor series development of the energy function associated with an evanescent wave. We then formulate another concept: the \textit{energy parallax}. It is defined mathematically in $\mathbf{Definition}$ $4$. Under some specific circumstances (e.g., the energy function exists), we show that the variations of energy locally in a predefined system, should lead to include additional solutions in the energy spaces with higher order $v$ (in $\mathbb{Z}^+$).  The second example is based on the local variations of EM energy density, which allows to define  waves which are first order derivative of  the EM field. This example is further explored in $\bold{Appendix}$ $IV$. Finally, the last example is the derivation of the Woodward effect with some strong hypothesis in order to include the EM energy density in the specific case of asymmetric EM cavity. We introduce in the Woodward effect, the first and second order derivative of the EM energy density, which can be interpreted such as a theoretical definition of an Electromagnetic and Gravitational coupling. 
\subsection*{Acknowledgements}
The author would like to acknowledge the important discussions with Dr. Jos\'e Rodal and Prof. Heidi Fearn (California State University Fullerton, physics department) on the Woodward effect and its derivation from general relativity. These discussions also helped to improve the manuscript. \textcolor{black}{In addition, we thank Prof. Paul Jolissaint (Universit\'e de Neuch$\hat{a}$tel, Institute of Mathematics) for his kind advices on the Sobolev spaces, and Prof. Marc Troyanov together with Dr. Luigi Provenzano (Ecole Polytechnique F\'ed\'erale de Lausanne, Institute of Mathematics) for their lecture of the manuscript and  corrections. Finally, we acknowledge the constructive comments from the anonymous reviewers improving this work.}
\subsection*{Appendix I: Generalities on Sobolev Spaces }
%
%
%
\textcolor{black}{A Sobolev space is a vector space of functions equipped with a norm that is a combination of $L^p$-norms of the function itself and its derivatives up to a given order. Intuitively, a Sobolev space is a space of functions with sufficiently many derivatives for some application domain, such as partial differential equations, and equipped with a norm that measures both the size and regularity of a function. Sobolev spaces are named after the Russian mathematician Sergei Sobolev.}
\vspace{0.5em}
\\$\bold{Definition}$ $I.1$ \textit{\cite{SobolevB}}:  Let  $\Omega \subseteq$ $\mathbb{R}^m$ ($m$ $\in$ $\mathbb{N}^+$) be open. The Sobolev space $W^{k,p}(\Omega)$ ($k \in \mathbb{N}$, $p \in [1, \infty]$) is defined as:
\begin{equation}
W^{k,p}(\Omega) =\{f \in L^p(\Omega)\lvert D^{\alpha} f \in L^p(\Omega), \forall \abs{\alpha} \leq k \}
\end{equation}
with $D^{\alpha} f$ the $\alpha$-th partial derivative in multi index notation, $\textcolor{black}{D^{\alpha} f = \frac{\partial^{\abs{\alpha}} f}{\partial t_1^{\alpha_1} \hdots \partial t_n^{\alpha_n}}}$.  The Sobolev space $W^{k,p}(\Omega)$ is the space of all locally integrable functions $f$ in $\Omega$ such as their partial derivatives $D^{\alpha} f$ exist in the weak sense for all multi index $\alpha \leq k$ and belongs to  $L^p(\Omega)$ (i.e. $\norm{f}_{L^p} < \infty$) (\cite{SobolevC}, chap. $5$). If $f$ lies in $W^{k,p}(\Omega)$, we define the $W^{k,p}$ norm of $f$ by the formula
\begin{equation}
\norm{f}_{W^{k,p}(\Omega)} = \sum_{|\alpha| \leq k} \norm{D^{\alpha} f}_{L^p(\Omega)}
\end{equation}
\\ Now, let us introduce the Fourier transform $\mathcal{F}: L^1(\mathbb{R}^m) \rightarrow \mathcal{C}_b (\mathbb{R}^m) $ as in \cite{SobolevA} 
\begin{equation}
\mathcal{F}(f) = \int_\Omega f(x) e^{-ix.\xi}dx = \mathcal{F}(f)(\xi)
\end{equation}
Here $\mathcal{C}_b (\mathbb{R}^m) $ is the space of bounded and continuous functions in $\mathbb{R}^m$ \cite{Folland}. Note that $.$ is the scalar product (with $x$ and $\xi$ in  $\mathbb{R}^m$). One can then define the Sobolev spaces for $\Omega = \mathbb{R}^m$,   $W^{k,p}(\mathbb{R}^m)$ using the Bessel potentials and the Fourier transform such as  \cite{SobolevB} or \cite{Folland} (chap. 9) :
\begin{equation}\label{HilbertSobolevdef}
W^{k,p}(\mathbb{R}^m) = H^{k,p}(\mathbb{R}^m) := \{f \in L^p(\mathbb{R}^m)\lvert \mathcal{F}^{-1}\big [ \big (1+ \abs{\xi}^2)^{k/2}\mathcal{F}(f) ] \in L^p(\mathbb{R}^m) \}
\end{equation}
The Bessel potential spaces are defined when replacing $k$ by any real number $s$. They are Banach spaces and, for the special case $p=2$, Hilbert spaces . 
Now, one can state an important result with Sobolev spaces   \cite{SobolevB}
\vspace{0.5em}\\$\bold{Theorem}$ $I.1$: $W^{k,p}(\mathbb{R}^m) \subseteq W^{l,q}(\mathbb{R}^m)$, whenever  $k > l \geq 0$ and $1\leq p <q <\infty$ are such that $(k-l)p<m$ 
\begin{proof}
\textcolor{black}{The proof of this theorem is rather long and technically delicate which is not our focus. Readers interested in this matter should refer to} \cite{SobolevB} \cite{SobolevC} (chap. 5) 
\end{proof}
%
%
%
%
\subsection*{Appendix II: Possible interpretation of the Energy Parallax in modern physics}
In Section \ref{Multiplicity}, we define mathematically the notion of \textit{multiplicity of solutions} for a given PDE. Through the various examples in Section \ref{Applications},  we define the concept of \textit{energy parallax}. The general meaning is that additional solutions should be taken into account when varying the amount of energy. Those solutions should be defined based on the associated energy spaces (e.g., $\mathbf{E}_p$ ($p$ $\in$ $\mathbb{Z}^+$). Now, if we replace this concept in modern physics, what is the meaning behind it?
\\ In modern physics, \textit{Energy} is a global concept across the whole science. The definition varies with for example kinetic energy and potential energy in classical mechanics. It relates respectively to the object's movement through space and function of its position  within a field \cite{Feynman}. Chemical energy can be defined  broadly such as the electrical potential energy among atoms and molecules. In quantum mechanics, energy is defined in terms of energy operators (e.g., Hamiltonian) as a time derivative of the work function. It allows to define particles at nominated energy levels associated with an EM waves emitted at frequencies defined by the Planck's relation. In General Relativity, energy results from the product of a varying mass and the square of the speed of light. Energy can describe the behavior of a system of two particles (and more) . For example, the electron-positron annihilation in which rest mass (invariant mass) is destroyed. At the opposite, the inverse process (creator) in which the rest mass of the particle is created from energy of two (or more) annihilating photons \cite{Encyclopedia}.
\\ \textit{Energy parallax} is here  defined such as the concept of using additional wave functions. For example in  Section \ref{Woodwardsection} increasing the higher  order derivatives of the EM energy density leads to the consideration of additional waves. The energy parallax concept can then help us to state that those additional waves are additional excited photons that we must take into account to vary the EM energy density. 
\subsection*{Appendix III: Discussion  on the possible relationship between the energy spaces $\mathbf{M}_{p}^{v}$ and $\mathbf{M}_{p+1}^{v-1}$}
%
%
This section follows the development in Section \ref{energyspace} and especially $\bold{Properties}$ $2$.
First, $\forall p \in \mathbb{Z}^+$, $ \mathbf{M}_{p+1}^{v} \bigcap \mathbf{M}_{p}^{v} \neq \{\emptyset\}$, because $0 \in \mathbf{S}^{-}(\mathbb{R}^m)$, and ($\forall p  \in \mathbb{Z}^+$, $v$ $ \in \mathbb{Z}^+$ ) $0 \in \mathbf{M}_p^v$. Thus, $0$ $\in  \mathbf{M}_{p+1}^{v} \bigcap \mathbf{M}_{p}^{v}$.
\\ To recall $\bold{Definition}$ $2$ and $\bold{Lemma}$ $2$, $\partial_i^v ([[f]^{p}]_1^+)^n $ can be decomposed with the family of energy operators $([[.]^{p+1}]_k^+)_{k\in \mathbb{Z}}$ ($\forall n \in \mathbb{Z}^+-\{0,1\}$, $i \in [1,...,m]$, $p\in \mathbb{Z}^+$, $v \in \mathbb{Z}^+-\{0\}$). Thus, one can write ($l<v$):
%
%
\begin{equation}
\partial_i^v ([[f]^{p}]_1^+)^n = \sum_{j=0}^{v-1} \big(_{j}^{v-1} \big) \partial_i^{v-1-j} ([[f]^{p}]_1^+)^{n-l} \sum_{u=-N_j}^{N_j} C_u [[\partial_i^{\alpha_u}f]^{p+1}]_1^+ 
\end{equation}
Thus, for $n>1$, $\bold{Lemma}$ $2$ allows to state that  $ \mathbf{M}_{p+1}^{v} \bigcap \mathbf{M}_{p}^{v} = \{\mathbf{M}_{p}^{v, n>1}\}$, with $\mathbf{M}_{p}^{v, n>1}$ the subspace of $\mathbf{E}_p$, but restricted for $n \in \mathbb{Z}^+$ and $n>1$.
%
\vspace{0.5em}
\\Furthermore, let us define the space $\mathbf{s}_p^{-,*}(\mathbb{R}^m)$ :
\begin{eqnarray}
\mathbf{s}_p^{-,*}(\mathbb{R}^m) &=& \{ f \in  \mathbf{S}_p^{-}(\mathbb{R}^m) | f \notin  \cup_{i\in [1,...,m]} \big ( \cup_{k \in \mathbb{Z}} Ker([[f]^p]_{k,i}^+) \big)\} \nonumber \\
\end{eqnarray}
Note that    $\mathbf{M}_p^\infty \not \subset  \mathbf{s}_p^{-,*}(\mathbb{R}^m)$, but the bump functions \cite{stevenG} are included in $\mathbf{s}_p^{-,*}(\mathbb{R}^m)$. We can also recall the discussion on $n=1$ in \cite{JPMontillet2017} and \cite{JPMontillet2014}, with the definition
\begin{equation}
\{\forall f \in \mathbf{s}_p^{-,*}(\mathbb{R}^m), \hspace{0.5em} p \in \mathbb{Z}^+ |\exists g \in \mathbf{S}^{-}(\mathbb{R}^m),  \hspace{0.5em} g=\frac{1}{\big ([[f]^p]_1^+\big )^n}, \hspace{0.5em} \forall n \in \mathbb{Z}^+, n>1\} 
\end{equation}
%
%
On can also state that $\partial_i^k \big ([[f]^p]_1^+\big ) = \partial_i^k \bigg (\frac{\big ([[f]^p]_1^+\big )^3}{\big ([[f]^p]_1^+\big )^2}\bigg)$ ($k \in \mathbb{Z}^+$) and use the Leibniz's rule for derivations in order to expand the multiple derivatives or the decomposition stated in $\bold{Lemma}$ $2$. 
If we call $\mathbf{M}_p^{v,*}$ ($p \in \mathbb{Z}^+$), the subspaces of $\mathbf{s}_p^{-,*}(\mathbb{R}^m)$. For all $g_1$ in $\mathbf{M}_p^{v,*}$ can be written as a non linear sum of  $g_2$ in $\mathbf{M}_{p+1}^{v,*}$. Finally, we can conclude that $\mathbf{M}_p^{v,*} \subsetneq \mathbf{M}_{p+1}^{v,*}$. With the specific extension of $\mathbf{M}_p^{v}$ to the case $n=1$, we can also conclude $\mathbf{M}_p^{v} \subsetneq \mathbf{M}_{p+1}^{v}$.
In addition, $\mathbf{s}_p^{-,*}(\mathbb{R}^m) \subsetneq \mathbf{s}_{p+1}^{-,*}(\mathbb{R}^m)$ by definition.
\subsection*{ Appendix IV: Consequences in terms of EM theory}
We are taking the example of the variation of EM energy density inside a copper wall due to planar waves reflecting and refracting on it \cite{Petit}. To recall Section \ref{sectionEMvariation}, the EM field is now including  ($\vec{E}$, $\vec{\delta E}$) and ($\vec{B}$, $\vec{\delta B}$),  contribution of the subspaces $\bold{M}_0^0$ and $\bold{M}_0^1$ respectively when using the concept of multiplicity of the solutions (i.e. $\bold{Theorem}$ $3$) for the higher order derivatives of the energy density (see \eqref{dweuqtion}). We call   the total EM field $\vec{E}_{tot}$ and $\vec{B}_{tot}$ inside the copper plate (skin layer) with associated permittivity $\epsilon$ and permeability $\mu$. They are solutions of the Maxwell equations:
\begin{equation}
    \left.\begin{array}{l}
       \textcolor{black}{\div} \vec{E}_{tot} = \frac{\rho_{tot}}{\epsilon},\\
        \textcolor{black}{\curl} \vec{E}_{tot}  = - \partial_t \vec{B}_{tot}, \\ 
        \textcolor{black}{\div} \vec{B}_{tot}   = 0,\\
        \textcolor{black}{\curl} \vec{B}_{tot}  = \mu \epsilon  \partial_t \vec{E}_{tot} + \mu \vec{j},
    \end{array}
    \right\} 
\end{equation}
with the principle of charge conservation: 
\begin{equation}
\partial_t \rho_{tot} + \textcolor{black}{\div} \vec{j} =0
\end{equation}
Now, the variation of energy density  \eqref{dweuqtion} together with the equation of charge conservation is formulated such as:
\begin{eqnarray}\label{energyDensityformula3}
\frac{d u}{dt} + \textcolor{black}{\div} \vec{P}_{tot} &=& \vec{j}.\vec{E}_{tot} \nonumber\\
\end{eqnarray}
$\vec{P}_{tot}  = \frac{\vec{E}_{tot}\times\vec{B}_{tot}}{\mu}$ is the Poynting vector. Now, writing $\vec{E}_{tot} = \vec{E} +\vec{\delta E}$, $\vec{B}_{tot} = \vec{B} +\vec{\delta B}$ and $\delta$ is the first derivative in time ($\partial_t$) (i.e. solutions in $\bold{M}_0^1$), then  following \cite{Petit}
\begin{eqnarray}
(\vec{E} +\vec{\partial_t E}).\vec{j} &=& (\vec{E} +\vec{\partial_t E}) . [\frac{1}{\mu} \textcolor{black}{\curl} \hspace{0.5em} (\vec{B} +\vec{\partial_t B}) -\epsilon \partial_t (\vec{E} +\vec{\partial_t E}) ] \nonumber \\
\end{eqnarray}   
using the equalities $\textcolor{black}{\div} \hspace{0.5em} (\vec{E}\times\vec{B}) = \vec{B}.\textcolor{black}{\curl}\vec{E} -\vec{E}.\textcolor{red}{\curl}\vec{B}$ and the Maxwell equation $\textcolor{black}{\curl}\vec{E} = -\partial_t \vec{B}$, $\textcolor{black}{\curl}\vec{\partial_t E} = -\partial^2_t \vec{B}$ the previous equation blackuces to:
\begin{eqnarray}\label{equalast}
\vec{E}. \vec{j} + \textcolor{black}{\div} \hspace{0.5em} (\frac{\vec{E}\times\vec{B}}{\mu}) + \partial_t  u + & & \nonumber \\
\vec{\partial_t E}. \vec{j} + \textcolor{black}{\div} \hspace{0.5em} (\frac{\partial_t\vec{E}\times\partial_t\vec{B}}{\mu}) + \partial_t^2  u + && \nonumber \\
 \textcolor{black}{\div} \hspace{0.5em} (\frac{\partial_t\vec{E}\times\vec{B}}{\mu})+  \textcolor{black}{\div} \hspace{0.5em} (\frac{\vec{E}\times \partial_t \vec{B}}{\mu}) + \frac{\partial \vec{B}.\partial \vec{B}}{\mu} +\epsilon \partial_t \vec{E}.\partial_t \vec{E} &=& 0
\end{eqnarray}
We can separate in three groups, 
\begin{equation*}
    \left.\begin{array}{l}
   \partial_t u+   \textcolor{black}{\div} \hspace{0.5em} (\frac{\vec{E}\times\vec{B}}{\mu})= - \vec{j}.\vec{E}\nonumber\\
   \partial_t^2 u +  \textcolor{black}{\div} \hspace{0.5em} (\frac{\partial_t\vec{E}\times\vec{B}}{\mu})+  \textcolor{black}{\div} \hspace{0.5em} (\frac{\vec{E}\times\partial_t \vec{B}}{\mu}) = -\vec{j}.\vec{\partial_t E} \\
  \textcolor{black}{\div} \hspace{0.5em} (\frac{\partial_t\vec{E}\times\partial_t\vec{B}}{\mu})   =- \frac{\partial_t \vec{B}.\partial_t \vec{B}}{\mu} -\epsilon \partial_t \vec{E}.\partial_t \vec{E}
    \end{array}
    \right\} 
\end{equation*} 
The Poynting vector is defined as $\vec{P} = \frac{\vec{E}\times\vec{B}}{\mu}$ and its derivative  $ \partial_t \vec{P} = \frac{\partial_t\vec{E}\times\vec{B}}{\mu}+\frac{\vec{E}\times\partial_t \vec{B}}{\mu}$. Thus, the second order term of the energy density is the contribution of the EM field generated by $\vec{\partial_t E}$ and $\vec{\partial_t B}$ is:
\begin{equation*}
    \left.\begin{array}{l}
   \partial_t u+  \textcolor{black}{\div} \vec{P} = -\vec{j}.\vec{E}\nonumber\\
   \partial_t^2 u  + \textcolor{black}{\div}  \hspace{0.5em}(\partial_t \vec{ P})  = -\vec{j}.\vec{\partial_t E} \nonumber\\
   \textcolor{black}{\div} \hspace{0.5em} (\frac{\partial_t\vec{E}\times\partial_t\vec{B}}{\mu})   =- \frac{\partial_t \vec{B}.\partial_t \vec{B}}{\mu} -\epsilon_0 \partial_t \vec{E}.\partial_t \vec{E}
    \end{array}
    \right\} 
\end{equation*}
The last line is the contribution from only the fields $\partial_t\vec{E}$ and $\partial_t\vec{B}$. 

Finally, the creation of the wave defined by the EM field   ($\partial_t\vec{E}$, $\partial_t\vec{B}$) means that some material properties may allow to create two type of EM waves namely ($\vec{E}$, $\vec{B}$) and ($\partial_t\vec{E}$, $\partial_t\vec{B}$).
%
%
%


\end{document}